\documentclass[aps,pra,10pt,twocolumn,superscriptaddress,longbibliography,floatfix]{revtex4-2}
\usepackage[utf8]{inputenc}
\usepackage[T1]{fontenc}
\usepackage{amsmath,amssymb,bm}
\usepackage{graphicx}
\usepackage{booktabs}
\usepackage{xcolor}
\usepackage[colorlinks=true,linkcolor=blue,citecolor=blue,urlcolor=blue]{hyperref}

\newcommand{\pc}{p_{\mathrm{c}}}
\newcommand{\ps}{p^{*}}
\newcommand{\Dmin}{\Delta_{\min}}
\newcommand{\dH}{d_{\mathrm{H}}}
\newcommand{\xc}{x_{\mathrm{c}}^{2}}
\newcommand{\bSB}{bSB$^{*}$}
\newcommand{\dSB}{dSB$^{*}$}

\begin{document}
\title{Revisiting the quantum origins of simulated bifurcation through configuration crossings}
\author{Yaocheng Chen}
\affiliation{Instituto de Astronomia, Geof\'{\i}sica e Ci\^encias Atmosf\'ericas, Universidade de S\~ao Paulo, S\~ao Paulo, SP, Brazil}
\author{Jiapei Zhuang}
\email{jpzhuangphy@gmail.com}
\affiliation{Quantum Science Center of Guangdong-Hong Kong-Macao Greater Bay Area, Shenzhen 518045, China}

\date{\today}
\begin{abstract}
Simulated bifurcation (SB) originated from quantum bifurcation in Kerr-parametric-oscillator (KPO) networks, yet this shared lineage does not imply shared dynamical bottlenecks on hard Ising instances. We examine this connection in closed KPO networks by varying an effective semiclassical parameter $h$ at fixed classical drift, linking classical configuration branches, quantum spectra, and finite-time evolution on the same instances while controlling for initial fluctuations. Across a four-spin test ensemble, classical configuration crossings correspond directly to quantum avoided crossings: extrapolated minimum-gap positions ($h\to 0$) align with classical crossing points, small-$h$ gaps predominantly follow exponential suppression whose effective action correlates with configuration separation, and a Landau--Zener estimate captures the principal finite-time quantum failures. Changing oscillator detunings at fixed Ising couplings yields consistent directional support: removing a crossing widens the minimum gap in all 40 affected instances (median ratio 1.73), whereas inducing one narrows it in all 26 (median ratio 0.63), with group-averaged classical success rates shifting correspondingly. By contrast, tuned ballistic SB removes the quantum success-rate advantage by modifying the dynamical rules, and an equal-duration hold inside a uniform-detuning Ising window substantially improves the success of a KPO-parameter ballistic model, narrowing the gap to tuned ballistic SB. These results separate tunneling bottlenecks in the parent oscillator model from the mechanisms of classical SB algorithms.
\end{abstract}
\maketitle
\section{Introduction}
Simulated bifurcation (SB) originated from a quantum-adiabatic proposal for Ising optimization in networks of two-photon-driven Kerr parametric oscillators (KPOs), where field-amplitude signs encode low-energy spin configurations~\cite{Goto2016SciRep}. Translating this picture into classical Hamiltonian dynamics yielded the original adiabatic SB algorithm~\cite{Goto2019SciAdv}, which has since branched into a rich family of classical solvers: ballistic and discrete formulations~\cite{Goto2021SciAdv}, heating-assisted and globally guided dynamics~\cite{Kanao2022CommunPhys,Xiao2026PRApplied}, replica-coupled extensions~\cite{Pawlowski2026arXiv}, and discrete-phase or native Potts generalizations~\cite{Chen2026ClockDSB,Zhuang2026PdSB}. Well beyond the original oscillator setting, SB variants have also been deployed in domains ranging from FPGA-based real-time trading~\cite{Tatsumura2023IEEEAccess} and molecular docking~\cite{Li2024JCTC} to quantum error decoding~\cite{Zhuang2026QECSB}. Because these algorithms modify the forces and boundary rules of the parent oscillator model, a shared Ising objective and historical lineage do not imply shared dynamical bottlenecks. This raises a central physical question: which optimization difficulties of the parent KPO network stem from configuration competition and quantum tunneling, and how do those difficulties transform when the classical dynamics are altered?

Answering this question requires distinguishing branch formation from branch competition across both classical and quantum regimes. On the classical side, continuous amplitudes introduce phenomena absent from the binary Ising landscape: amplitude heterogeneity can distort the analog-to-discrete mapping~\cite{Leleu2019PRL}, while retarded bifurcation and local minima govern early dynamical trapping in coherent Ising machines and related continuous solvers~\cite{Wang2023NatCommun,Sato2022arXiv}. On the quantum side, parametrically driven Kerr oscillators exhibit nontrivial tunneling physics, including detuning-dependent interference oscillations~\cite{Marthaler2007PRA} and level-crossing structures organized by classical phase-space geometry~\cite{Prado2023PRA,Prado2025}, while excited-state KPO protocols exploit nonadiabatic transitions near small gaps~\cite{GotoKanao2020CommunPhys}. Whereas prior studies largely emphasize bifurcation onset or single-oscillator tunneling, our focus here is the energy crossing of already existing configuration branches in coupled networks and its direct manifestation as a quantum spectral bottleneck.

While KPO annealing admits multiple encoding architectures~\cite{Puri2017NatCommun}, we ground our analysis in the two-body network of Goto~\cite{Goto2016SciRep} and introduce a semiclassical family that varies an effective Planck constant $h$ while holding the classical drift fixed. This construction places classical configuration branches, quantum energy spectra, and finite-time dynamics on a common footing for the same Ising instances. We focus primarily on four-spin networks, where all $2^{n-1}$ configuration branches can be tracked exhaustively alongside resolved small-$h$ spectra and direct Schr\"odinger propagation, and use six-spin instances to test how these mechanisms extend to a larger network.

Our analysis establishes three main results. First, classical configuration crossings organize the principal quantum spectral bottlenecks and finite-time failures of the KPO network, as demonstrated by extrapolated minimum-gap positions, finite-range exponential gap suppression, and instance-level agreement between Landau--Zener estimates and direct quantum evolution. Second, matching initial fluctuation scales separates early amplification delay from persistent crossing-induced difficulty, and paired detuning interventions at fixed Ising couplings provide consistent directional support by shifting quantum gaps and group-averaged classical success rates in tandem. Third, comparing against classical SB variants clarifies the algorithmic boundary of this shared origin: tuned ballistic SB eliminates the quantum success-rate advantage on the four-spin test set through altered dynamical rules, while equal-duration hold tests show that evolving inside a uniform-detuning Ising window substantially improves the success of a KPO-parameter ballistic counterpart and narrows the gap to tuned ballistic SB.

The remainder of the paper is organized as follows. Section~\ref{sec:methods} defines the KPO model families, classical counterparts, SB algorithms, and evaluation protocols. Section~\ref{sec:classical} isolates residual classical failures after controlling for early amplification and initial fluctuations. Section~\ref{sec:quantum} connects configuration crossings to quantum spectra and finite-time dynamics. Sections~\ref{sec:intervention} and \ref{sec:sb} test the detuning intervention, analyze classical SB alternatives, and examine six-spin extensions. The concluding section discusses the physical implications and scope of these findings.

\section{Model and comparison protocols}
\label{sec:methods}

\begin{figure}[t]
\centering
\includegraphics[width=\columnwidth]{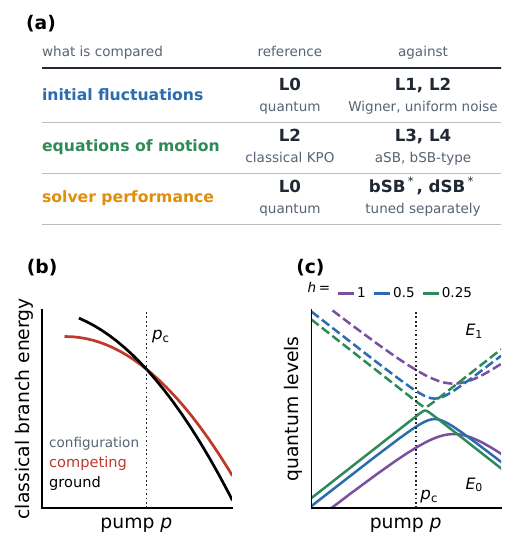}
\caption{Comparison framework and schematic crossing correspondence. (a) The three comparisons made on the same Ising instances: initial fluctuations (L0 against L1 and L2), equations of motion (L2 against L3 and L4), and performance as a solver (L0 against the separately tuned \bSB{} and \dSB{}). L1 and L2 share the classical KPO drift with Wigner and uniform initialization, respectively; L3 and L4 retain KPO parameters under modified dynamics, whereas tuned SB uses its own normalization, parameters, and step budget. Stars denote training-selected settings. (b,c) Conceptual schematics of (b) a competing branch exchanging classical energy order with the Ising ground configuration at $\pc$, and (c) the associated even-parity quantum avoided crossing at several values of $h$ (solid: $E_0$; dashed: $E_1$), showing the finite-$h$ shift of $\ps$ relative to $\pc$ (vertical dotted line).}
\label{fig:schematic}
\end{figure}

To relate classical configuration structure, quantum spectra, and final-time success on common Ising instances, we compare the closed KPO network against classical counterparts and independently tuned SB algorithms (Fig.~\ref{fig:schematic}). Below, we first define the KPO Hamiltonian, fixed-drift semiclassical scaling, classical and SB baselines, and evaluation protocols (Secs.~II\,A--II\,D), and then specify how configuration branches, quantum gaps, and Landau--Zener estimates are computed (Secs.~II\,E and II\,F).

\subsection{KPO network and schedule}

The KPO network encodes Ising variables in the signs of oscillator quadratures, while a time-dependent two-photon pump drives the evolution from vacuum. Following Goto's setting~\cite{Goto2016SciRep}, we consider $n$ coupled oscillators with Hamiltonian, in a frame rotating at half the pump frequency and with $\hbar=1$,

\begin{equation}
\begin{split}
\hat H(t)={}&\sum_{i=1}^{n}\left[\tilde\Delta_i\hat a_i^\dagger\hat a_i+\frac{Kh}{2}\hat a_i^{\dagger2}\hat a_i^2-\frac{p(t)}2(\hat a_i^2+\hat a_i^{\dagger2})\right]\\
&-\xi_0\sum_{i\ne j}J_{ij}\hat a_i^\dagger\hat a_j .
\end{split}
\label{eq:H}
\end{equation}

Here $\hat a_i$ annihilates an excitation in oscillator $i$, and the terms represent detuning, Kerr nonlinearity, two-photon driving, and inter-oscillator coupling, respectively. We use $K=1$ as the energy unit, $1/K$ as the time unit, and $\xi_0=0.5K$. The couplings satisfy $J_{ij}=J_{ji}$ and $J_{ii}=0$, defining the Ising objective $E(\boldsymbol{s})=-\tfrac12\sum_{ij}J_{ij}s_is_j$. Unless stated otherwise, the pump ramps linearly as $p(t)=p_ft/T$, with $p_f=7K$ and $T=700/K$.

We define the base detunings $\Delta_i=\xi_0\sum_j|J_{ij}|$ and the positive-semidefinite matrix

\begin{equation}
M=\operatorname{diag}(\Delta)-\xi_0J.
\label{eq:M}
\end{equation}

The Hamiltonian detunings $\tilde\Delta_i$ are specified separately for the two model families below, separating the quantum initial-state condition from the linear stability of the associated classical flow.

Readout uses $s_i=\operatorname{sgn}(x_i)$. Quantum sign-pattern probabilities are evaluated in the truncated Fock basis using $\Theta(\hat x_i)$, where $\hat x_i=(\hat a_i+\hat a_i^\dagger)/\sqrt2$; the classical coordinate $x_i=\operatorname{Re}z_i$ below uses the same sign convention. Success is the total probability of all Ising ground-state sign patterns (including global sign reversal), estimated for classical dynamics by the corresponding trajectory frequency.

\subsection{Effective Planck constant}

To compare quantum spectra against a fixed classical configuration landscape, we introduce an effective semiclassical parameter $h$ via the rescaled operators $\hat z_i=\sqrt h\,\hat a_i$, which obey $[\hat z_i,\hat z_j^\dagger]=h\delta_{ij}$. Up to an additive constant, the Weyl symbol of $h\hat H$ is

\begin{equation}
\begin{split}
H_{\mathrm{cl}}(\boldsymbol z)={}&\sum_i\left[(\tilde\Delta_i-Kh)|z_i|^2+\frac K2|z_i|^4-\frac p2(z_i^2+z_i^{*2})\right]\\
&-\xi_0\sum_{i\ne j}J_{ij}z_i^*z_j .
\end{split}
\label{eq:Hcl}
\end{equation}

In the \emph{weyl} family we set $\tilde\Delta_i=\Delta_i+Kh$, which fixes the effective classical detuning at $\Delta_i$ and yields the classical flow

\begin{equation}
i\dot z_i=(\Delta_i+K|z_i|^2)z_i-pz_i^*-\xi_0\sum_jJ_{ij}z_j .
\label{eq:flow}
\end{equation}

This reproduces Goto's classical KPO equation~\cite{Goto2016SciRep} for every $h$, so varying $h$ changes the effective quantum scale while holding the classical drift fixed. In the \emph{goto} family, $\tilde\Delta_i=\Delta_i$: setting $h=1$ recovers Goto's original quantum Hamiltonian, whereas its Weyl classical flow carries detuning $\Delta_i-Kh$. The two families coincide as $h\to0$. Unless stated otherwise, results refer to the weyl family at $h=1$.

At $p=0$, the number-conserving quadratic part of $\hat H$ has matrix $M+KhI$ in the weyl family and $M$ in the goto family. Both are positive semidefinite, which together with the nonnegative Kerr term ($K,h>0$) ensures that the vacuum is a ground state in both families. Linearizing the weyl classical flow about the origin gives $\dot{\boldsymbol x}=(M+pI)\boldsymbol y$ and $\dot{\boldsymbol y}=-(M-pI)\boldsymbol x$, placing the first linear instability at $p=\lambda_1(M)$. In the goto Weyl flow the corresponding matrix is $M-KhI$, shifting this threshold.

In the reported calculations, the final photon-number scale is approximately $6.2/h$ per mode. Although effective spin models offer a complementary description of KPO annealing~\cite{Miyazaki2022PRA,Stenzel2026PRA}, retaining oscillator variables allows direct comparison between classical configuration branches and the quantum spectrum.

\subsection{Classical counterparts and SB algorithms}
\label{sec:layers}

To separate initialization effects from dynamical-rule changes, we compare the quantum reference (L0) with four classical KPO-parameter models (L1--L4) and two independently tuned SB baselines (\bSB{} and \dSB{}), as organized in Fig.~\ref{fig:schematic}(a).

L0 is direct Schr\"odinger evolution of Eq.~\eqref{eq:H} in the even total-parity sector of the truncated Fock space. L1 is the truncated Wigner approximation (TWA), evolving Eq.~\eqref{eq:flow} from vacuum Wigner initial conditions $\operatorname{Re}z_i,\operatorname{Im}z_i\sim\mathcal N(0,h/4)$~\cite{Polkovnikov2010AnnPhys}. L2 integrates the same classical drift with both initial quadratures drawn uniformly from $[-\varepsilon,\varepsilon]$; in addition to Goto's choice $\varepsilon=10^{-6}$, we scan $\varepsilon\in[10^{-8},1]$ and select $\varepsilon^{*}$ on a training set.

L3 (aSB~\cite{Goto2019SciAdv}) and L4 (a bSB-type model) retain the KPO parameters while altering the equations of motion:

\begin{equation}
\begin{aligned}
\mathrm{L3}:\quad &\dot y_i=-(Kx_i^2-p+\Delta_i)x_i+\xi_0\sum_jJ_{ij}x_j,\\
\mathrm{L4}:\quad &\dot y_i=-(\Delta_i-p)x_i+\xi_0\sum_jJ_{ij}x_j,\\
&\dot x_i=\Delta_i y_i\quad (\mathrm{L3,L4}),
\end{aligned}
\end{equation}

where $x_i$ carries the Ising sign and $y_i$ is the conjugate momentum. Compared with the full KPO flow, L3 simplifies the momentum coupling to $\dot x_i=\Delta_i y_i$ while keeping cubic confinement $-Kx_i^3$; L4 removes the cubic term and imposes inelastic walls at $x_w=\sqrt{p_f/K}$, resetting $x_i=x_w\operatorname{sgn}(x_i)$ and $y_i=0$ upon any overshoot $|x_i|>x_w$.

As algorithmic baselines beyond the KPO parameter setting, we implement bSB and dSB using the standard normalization of Ref.~\cite{Goto2021SciAdv}. Each step updates $y$ first and then $x$:

\begin{equation}
\begin{aligned}
y_i&\leftarrow y_i+\delta t\left[-(a_0-a_k)x_i+c_0\sum_jJ_{ij}f(x_j)\right],\\
x_i&\leftarrow x_i+\delta t\,a_0y_i,
\end{aligned}
\end{equation}

with $a_0=1$, linear ramp $a_k=(k+1)/N_{\mathrm{steps}}$ ($k=0,\ldots,N_{\mathrm{steps}}-1$), and $c_0=c/(\sigma_J\sqrt n)$, where $\sigma_J=[n(n-1)]^{-1/2}(\sum_{i\ne j}J_{ij}^{2})^{1/2}$ is the root-mean-square off-diagonal coupling. Here $f(x)=x$ for bSB and $f(x)=\operatorname{sgn}(x)$ for dSB, with inelastic walls at $|x_i|=1$. The parameters $c$, $\delta t$, $N_{\mathrm{steps}}$, and initial amplitude are chosen by training-set grid search (Appendix~\ref{app:hyper}), yielding the selected baselines \bSB{} and \dSB{} in Table~\ref{tab:baselines}.

L1 and L2 are integrated with fourth-order Runge--Kutta, while L3 and L4 use symplectic Euler (pump evaluated at the updated time, $y$ updated before $x$) with $\delta t=0.0025$. On paired step-refinement checks (G4 instances 0--9, $\delta t$ halved to $0.00125$), L2--L4 success probabilities change by at most $0.006$ per instance and $6\times10^{-4}$ in the mean (Appendix~\ref{app:conv}). Tuned \bSB{} and \dSB{} use their own algorithmic step sizes.

\subsection{Instances and comparison protocols}

Our evaluation separates held-out performance testing, targeted mechanism analyses, and six-spin transfer tests; full sample definitions and index ranges are summarized in Table~\ref{tab:samples} (Appendix~\ref{app:hyper}).

Following Goto~\cite{Goto2016SciRep}, the primary four-spin ensemble G4 comprises 1000 instances with independent off-diagonal couplings $J_{ij}$ ($i<j$) drawn uniformly from $[-1,1]$. Instances 0--299 form the held-out performance test set (containing 17 crossings). Training sets disjoint from the test set select the L2--L4 initial noise (instances 300--499) and the bSB/dSB hyperparameters (instances 300--999), each maximizing mean training success over its search grid (Appendix~\ref{app:hyper}). For the small-$h$ spectral analysis, we use a 60-instance subset within the test range comprising instances 0--49 plus all 17 test crossings (43 noncrossing controls). Additional noise and intervention subsets are specified where used.

To test transfer to a larger network, we generate 300 six-spin instances (G6) by the same procedure and evaluate all classical methods using the four-spin-selected parameters without retuning. Because the physical couplings are not renormalized with $n$, the mean base detuning increases from approximately $0.75K$ to $1.25K$ at fixed $p_f$. For G6, quantum success is estimated via Landau--Zener (LZ) rates on a 66-instance spectral subset (all 36 crossings plus 30 noncrossing controls); direct six-spin quantum propagation is not performed, and quantum--classical comparisons are restricted to common subsets.

Classical success probabilities are estimated from $R=1000$ trajectories per instance, and ensemble means weight instances equally. Because the physical KPO models use a fixed ramp $T=700/K$ whereas \bSB{} and \dSB{} use $10^4$ algorithmic steps ($\delta t=1$ and $\delta t=0.5$, respectively) trained over a larger search space, our comparisons evaluate final success rates under each method's stated protocol rather than equal-cost speedup or time-to-solution scaling~\cite{Ronnow2014Science}.

\begin{table*}[t]
\caption{Success-rate baselines under each method's stated protocol. KPO-parameter layers use the weyl family at $h=1$ with $T=700/K$; \bSB{} and \dSB{} use the algorithmic settings listed below (Appendix~\ref{app:hyper}). Parameters trained on four-spin sets are transferred to six spins without retuning. ``All'' averages over 300 test instances; ``Cross.'' averages over 17 four-spin or 36 six-spin crossings. The last column counts four-spin test instances with $P_Q-P_B>0.1$. Six-spin quantum results are LZ estimates on 66 selected instances (including all 36 crossings; Sec.~\ref{sec:sb}), so no 300-instance quantum mean is reported.}
\label{tab:baselines}
\begin{ruledtabular}
\begin{tabular}{lccccc}
 & \multicolumn{2}{c}{$n=4$} & \multicolumn{2}{c}{$n=6$} & \\
Baseline & all & cross. & all & cross. & $Q-B>0.1$\\
\hline
L2, $\varepsilon=10^{-6}$ (Goto) & 0.886 & 0.400 & 0.857 & 0.452 & 90\\
L2, $\varepsilon^{*}=0.03$ & 0.948 & 0.337 & 0.869 & 0.239 & 31\\
L3 (aSB), $\varepsilon^{*}=10^{-3}$ & 0.958 & 0.351 & 0.865 & 0.260 & 17\\
L4 (bSB-type), $\varepsilon^{*}=10^{-6}$ & 0.964 & 0.544 & 0.895 & 0.432 & 14\\
\bSB{} ($c=0.5$, $\delta t=1$, $10^4$ steps) & 1.000 & 1.000 & 0.967 & 0.972 & 0\\
\dSB{} ($c=0.5$, $\delta t=0.5$, $10^4$ steps) & 0.991 & 0.955 & 0.939 & 0.888 & 8\\
\hline
Quantum L0 & 0.996 & 0.927 & -- & 0.927 (LZ) & \\
\end{tabular}
\end{ruledtabular}
\end{table*}

\begin{figure*}[t]
\centering
\includegraphics[width=0.95\textwidth]{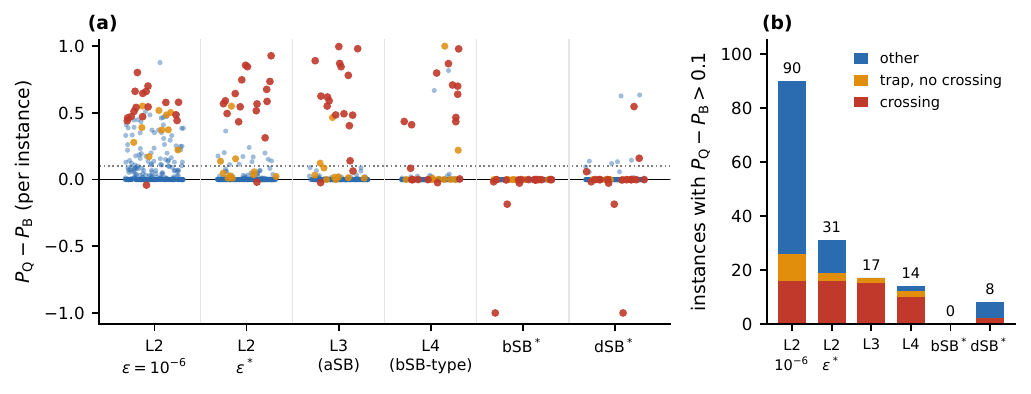}
\caption{Instance-resolved performance on the 300 four-spin test instances (G4 0--299). (a) Advantage $P_Q-P_B$ of the quantum reference over each baseline, one point per instance (red: crossing; orange: trap without crossing; blue: other; points spread horizontally within each column). The dotted line marks an advantage of 0.1. Negative values are instances on which the baseline does better, among them \#267, which \bSB{} and \dSB{} solve and the quantum reference does not. (b) Number of instances with $P_Q-P_B>0.1$ for each baseline, stacked by instance class. Stars denote training-selected parameters.}
\label{fig:who}
\end{figure*}

\subsection{Classical configuration branches}
\label{sec:branches_method}

Competition between Ising sign configurations is captured by the local-minimum branches of the classical energy along the pump ramp. For real amplitudes $\boldsymbol z=\boldsymbol x$, the weyl-family energy~\eqref{eq:Hcl} reduces to

\begin{equation}
V(\boldsymbol x;p)=\sum_i\left[(\Delta_i-p)x_i^2+\frac K2x_i^4\right]-\xi_0\sum_{i\ne j}J_{ij}x_ix_j.
\label{eq:V}
\end{equation}

Because the continuous amplitudes adjust as $p$ increases, competing sign configurations can exchange their ordering in $V$ even though the discrete Ising objective is fixed. Each sign pattern $\boldsymbol s$ defines an orthant $\operatorname{sgn}\boldsymbol x=\boldsymbol s$; identifying patterns related by global sign reversal leaves $2^{n-1}$ sign classes. A configuration branch tracks a local minimum of $V(\boldsymbol x;p)$ within its orthant as $p$ varies, with branch energy differences divided by $h$ giving the energy scale of $\hat H$.

Numerically, we trace all $2^{n-1}$ classes by damped Newton continuation from $p=p_f$ down to $p=0$ in steps of 0.005, terminating a branch if convergence, sign consistency, or local-minimum stability is lost. An instance is classified as a \emph{crossing} when a competing branch lies below all Ising ground-configuration branches over some pump interval before a ground branch becomes lowest at $p=\pc$ and remains lowest through $p_f$. Numerically, $\pc$ is recorded at the first grid point after the last occurrence of a lower competing branch.

At the recorded $\pc$, we characterize the separation between the competing and ground configurations by their Hamming distance $\dH$ (minimized over global reversal) and the ground-branch mean squared amplitude $\xc=n^{-1}\sum_i x_i^2$; the ratio $\xc/h$ estimates the coherent-state photon number per mode, and we refer to $\dH\xc$ as the classical distance between the two configurations. Instances whose Ising ground configuration is still not the lowest branch at $p_f$ are labeled \emph{non-ground-terminating} (NGT) and kept separate from ordinary noncrossing controls. Finally, an \emph{eigenvector trap} denotes an instance whose lowest-$M$-eigenvector sign pattern is not an Ising ground state~\cite{Goto2016SciRep}: thus the trap label reflects the first unstable linear mode, whereas the crossing label reflects the nonlinear branch ordering over the ramp.

\subsection{Quantum dynamics and gaps}
\label{sec:quantum_method}

Direct quantum propagation and spectral calculations test whether the classical configuration structure governs finite-time quantum failures. Direct propagation evolves the vacuum under Eq.~\eqref{eq:H} in a Fock basis truncated to occupations $0,\ldots,n_{\max}$ per mode ($d=n_{\max}+1$) using a fourth-order commutator-free Magnus integrator~\cite{Alvermann2011JCP} with adaptive Lanczos exponentials.

Because $\hat H(p)$ conserves total photon-number parity and the vacuum has even parity, evolution remains within the even-parity sector. Let $E_0(p)$ and $E_1(p)$ denote the two lowest even-parity eigenvalues; the instantaneous gap is $g(p)=E_1(p)-E_0(p)$, with minimum $\Dmin$ at pump $\ps$. We track the four lowest even-parity eigenvalues along a pump grid via block-wise exact diagonalization at $p=0$ and LOBPCG continuation at $p>0$~\cite{Knyazev2001SIAM}, refining the deepest candidate minima of $g(p)$ with bounded Brent search.

Near an isolated avoided crossing, the local gap follows the two-level form~\cite{Zener1932}

\begin{equation*}
g(p)^2\simeq\Dmin^2+v^2(p-\ps)^2,
\end{equation*}

where $v^2=\Dmin\,g''(\ps)$ is set by the curvature of the gap at its minimum, estimated as the median of $[g(p)^2-\Dmin^2]/(p-\ps)^2$ at $p-\ps=\pm10^{-3},\pm3\times10^{-3},\pm10^{-2},\pm3\times10^{-2}$; for a hyperbolic gap it equals the asymptotic slope of the diabatic levels. For the linear ramp with velocity $p_f/T$, the corresponding Landau--Zener nonadiabatic excitation probability is

\begin{equation}
P_{\mathrm{LZ}}=\exp\left[-\frac{\pi\Dmin^2T}{2vp_f}\right],
\label{eq:LZ}
\end{equation}

assuming a locally isolated avoided crossing entered predominantly in the lower adiabatic state. Its accuracy as a predictor of final Ising failure is tested directly against four-spin quantum propagation.

As a truncation diagnostic, we monitor the population in the two highest retained Fock levels, targeting values below $\sim 10^{-4}$. Reported four-spin calculations use $n_{\max}=22$ at $h=1$ up to $n_{\max}=72$ at $h=0.12$ (even-sector dimensions $\sim 1.4\times10^5$ to $1.4\times10^7$), with custom GPU kernels for small-$h$ spectra. Most propagation runs were performed on GPUs, which propagate batches of instances with the same integrator; on test systems the GPU and CPU implementations agree to $3\times10^{-14}$. Truncation diagnostics and solver checks are detailed in Appendices~\ref{app:conv} and \ref{app:gap}.

\begin{figure*}[t]
\centering
\includegraphics[width=0.95\textwidth]{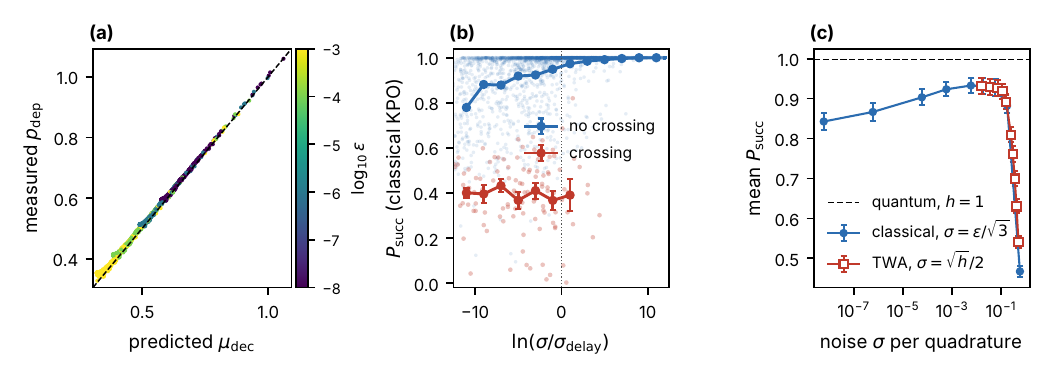}
\caption{Early amplification and initialization dependence. (a) Predicted versus trajectory-median measured departure pump across 2000 L2 conditions on G4 0--499 ($\varepsilon\leq10^{-3}$, departure fraction $>0.99$); color encodes $\log_{10}\varepsilon$. (b) Final L2 success across 2500 conditions ($\varepsilon\leq10^{-2}$) on the same 500 instances versus $\ln(\sigma/\sigma_{\mathrm{delay}})$, where positive values predict departure before the second linear instability; faint points are single conditions, connected symbols are class means in bins of width 2 with standard errors (the end bins collect all values beyond $\pm10$). (c) Mean success of L2 (filled circles) and TWA (open squares) on G4 0--99 versus initial quadrature standard deviation, alongside the $h=1$ quantum reference (dashed line); error bars show standard errors across instances.}
\label{fig:classical}
\end{figure*}

\section{Classical structure and residual failures}
\label{sec:classical}

To separate aggregate optimization performance from instance-level bottlenecks, we first compare the quantum reference against classical KPO counterparts and tuned SB baselines on the four-spin test set (Sec.~III\,A). We then isolate early amplification delay and initialization effects via noise scans and matched-fluctuation TWA comparisons (Secs.~III\,B and III\,C), identifying the residual failures examined spectrally in Sec.~\ref{sec:quantum}.

\subsection{From aggregate success to difficult instances}
On the four-spin test set, the quantum reference outperforms the classical KPO-parameter models, whereas \bSB{} attains a higher mean success probability. Figure~\ref{fig:who} resolves these comparisons by instance, and Table~\ref{tab:baselines} summarizes the ensemble and crossing-subset means.

The 300-instance four-spin test set contains 17 crossings, 10 eigenvector traps without crossing, and 273 other instances. Quantum mean success is 0.9958 overall and 0.9272 on crossings. Relative to L2 at $\varepsilon=10^{-6}$, quantum success exceeds classical success by more than 0.1 on 90 instances; the corresponding counts drop to 31 for noise-tuned L2, 17 for L3, 14 for L4, and zero for \bSB{}. Most of the 90 instances against L2 at $\varepsilon=10^{-6}$ are noncrossing instances (64 others and 10 traps without crossing), which noise tuning recovers; the advantage that remains concentrates on the 17 crossings, which account for 16 of the 31 instances against noise-tuned L2, 15 of 17 against L3, and 10 of 14 against L4 [Fig.~\ref{fig:who}(b)]. Moreover, Fig.~\ref{fig:who}(a) shows that \bSB{} succeeds on instances where the quantum reference retains failures.

Noise tuning raises the L2 test-set mean from 0.886 to 0.948, but lowers its mean on the crossing subset from 0.400 to 0.337, showing that optimizing ensemble-wide mean success does not uniformly resolve this structurally difficult subset. On the 17 crossings, training-selected L2, L3, and L4 achieve mean success probabilities of 0.337, 0.351, and 0.544. Even their per-instance retrospective envelope (taking the highest success among the three layers for each instance) averages only 0.571, compared with 0.927 for the quantum reference, which exceeds this envelope by more than 0.1 on 10 of the 17 crossing instances (while a few noncrossing instances also retain failures).

Across the full 1000-instance G4 ensemble, eigenvector traps occur in 9.1\% of instances and account for 53 of the 54 crossing instances, leaving only one crossing outside the trap class. In all 54 weyl crossing instances, the ground-configuration branch already exists below $\pc$, confirming that the crossing exchanges the energy order of preexisting branches.

\subsection{Early amplification does not determine every final sign}
Small initial amplitudes require finite time to grow once the origin loses stability. For the weyl flow, mode $k$ becomes linearly unstable at $p=\lambda_k$ (an eigenvalue of $M$) with instantaneous growth rate $\sqrt{p^2-\lambda_k^2}$. Integrating this rate along the linear ramp $r=p_f/T$ yields the logarithmic amplification

\begin{equation}
\Lambda_k(p)=\frac1r\int_{\lambda_k}^{p}\sqrt{q^2-\lambda_k^2}\,dq,\qquad r=p_f/T.
\end{equation}

In this linear-growth approximation, an initial quadrature standard deviation $\sigma$ grows as $\sigma e^{\Lambda_1(p)}$; equating it to a departure threshold $\ell=0.1$ defines the predicted departure pump $\mu_{\mathrm{dec}}$ via $\Lambda_1(\mu_{\mathrm{dec}})=\ln(\ell/\sigma)$. Measured departure is the first passage of $\max_i|z_i|$ through $\ell$, summarized in Fig.~\ref{fig:classical}(a) by the trajectory median for each instance--noise condition.

On the 500 G4 instances 0--499, across the 2000 conditions with $\varepsilon\leq10^{-3}$ and more than 99\% of trajectories departing, the median measured-minus-predicted departure pump is $+0.008$, with 90\% of absolute deviations below 0.022 and a Spearman rank correlation of 0.997. Defining $\sigma_{\mathrm{delay}}=\ell\exp[-\Lambda_1(\lambda_2)]$, the condition $\ln(\sigma/\sigma_{\mathrm{delay}})>0$ predicts departure before the second mode becomes unstable (an instance-dependent timing threshold distinct from the training-selected $\varepsilon^{*}$).

Figure~\ref{fig:classical}(b) shows final L2 success over 2500 conditions ($\varepsilon\leq10^{-2}$) on the same 500 instances. Noncrossing instances generally achieve high success once departure precedes the second instability, whereas crossing-related failures persist across the scanned noise range.

A separate trajectory check on all 54 G4 crossing instances (200 trajectories per instance, $\varepsilon=10^{-6}$, configurations assigned from real amplitudes averaged over pump windows of width 0.01) further shows that dynamics remain active well past initial amplification. After a settling interval of 0.2 in pump beyond departure, about 84\% of trajectories undergo at least one subsequent configuration change, and ground-state membership at the end of this interval agrees with final success for only about 64\% of trajectories. Early sign selection therefore provides only partial information about the final outcome, showing that the branch crossing reflects the nonlinear energy landscape rather than a frozen early capture.

\begin{figure}[htbp]
\centering
\includegraphics[width=0.96\columnwidth]{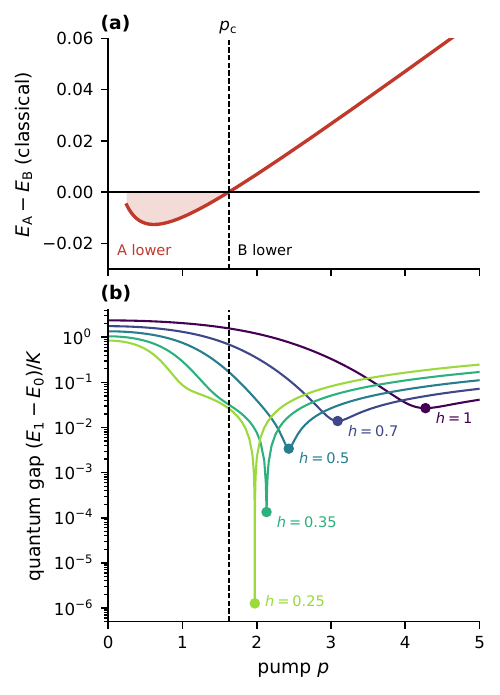}
\caption{Configuration competition and quantum bottleneck for weyl instance \#29. (a) Classical energy of the competing configuration A relative to the Ising ground configuration B, $E_{\rm A}-E_{\rm B}$, from Eq.~\eqref{eq:V}. A is lower below $\pc=1.625$ (shaded) and B above it (vertical dashed line). (b) Lowest even-parity quantum gap of $\hat H$ for $h=1$ to $0.25$; filled circles mark the minimum gap $\Dmin$ at $\ps$. As $h$ decreases, the minimum moves toward $\pc$ and deepens. Each curve joins the coarse $p$ grid with the points of the minimum search; at $h=0.18$ and $0.12$ the minimum lies below the $10^{-6}K$ resolution and is not drawn.}
\label{fig:anatomy}
\end{figure}

\subsection{Matching initial fluctuations}
For the weyl family, TWA (L1) and L2 share the same classical drift with Gaussian and uniform initial distributions whose per-quadrature standard deviations are $\sqrt h/2$ and $\varepsilon/\sqrt3$, respectively. Figure~\ref{fig:classical}(c) compares both ensembles on this common standard-deviation axis ($\varepsilon/\sqrt3=\sqrt h/2$) over G4 instances 0--99, varying $h$ in the TWA initialization while keeping the quantum reference fixed at $h=1$.

On this 100-instance set, TWA success rises from 0.541 at $h=1$ to 0.931 at $h=10^{-3}$, compared with 0.9979 for the quantum reference at $h=1$. The close overlap of the L2 and TWA curves on the matched noise-scale axis demonstrates that initial fluctuation amplitude accounts for most of their difference under the shared drift. At $h=1$ the TWA success differs from the quantum one by $0.46$ on average and by $0.66$ on the nine crossing instances of this set; since TWA follows L2 at equal noise and L2 fails on crossings at every noise level [Fig.~\ref{fig:classical}(b)], no initial noise brings either close to the quantum success there. The remaining gap to the quantum reference motivates the spectral analysis of difficult instances in Sec.~\ref{sec:quantum}.

\begin{figure*}[t]
\centering
\includegraphics[width=0.95\textwidth]{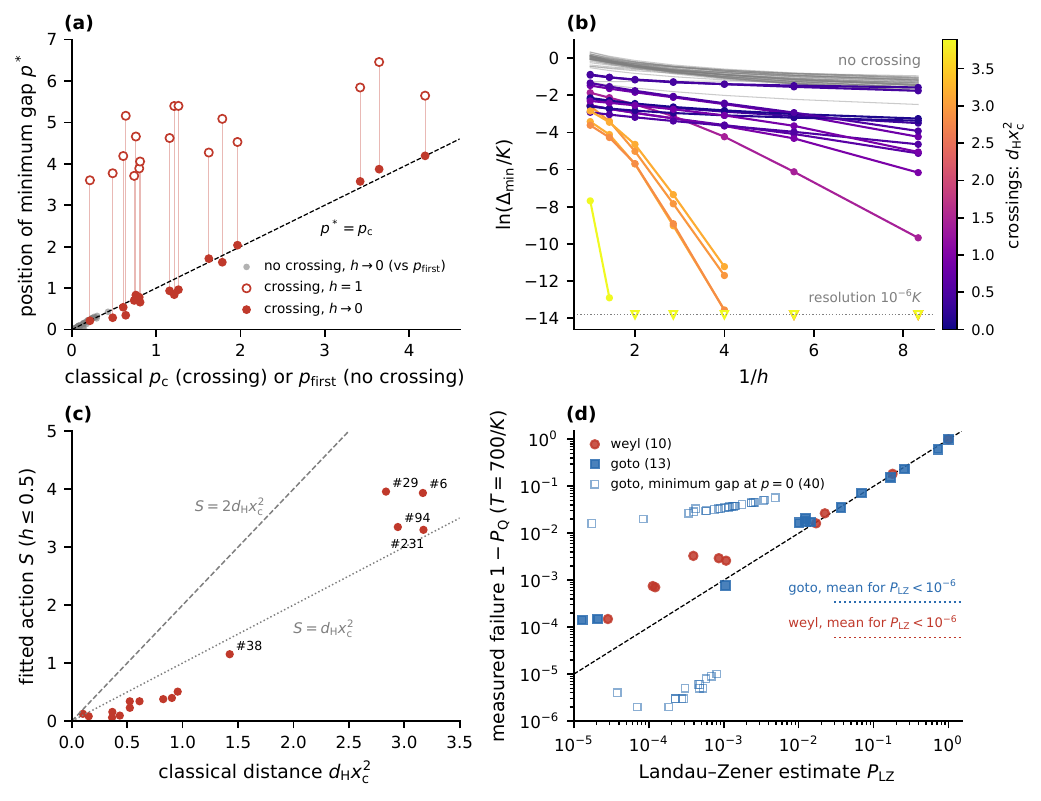}
\caption{Position correspondence, finite-range gap scaling, and dynamical validation. (a--c) Selected weyl spectral sample (17 crossings, 43 controls). (a) Position $\ps$ of the minimum gap against the classical reference, $\pc$ for crossings and $p_{\mathrm{first}}$ for controls. Open circles: $h=1$; filled circles: quadratic extrapolation to $h\to0$ over all seven $h$ values; dashed line: $\ps=\pc$. (b) $\ln(\Dmin/K)$ versus $1/h$; crossings colored by $\dH\xc$, controls gray; open triangles mark gaps below the $10^{-6}K$ resolution (excluded from the fits, which use $h\leq0.5$, $\Dmin>10^{-6}K$, and $\geq3$ points). (c) Fitted effective action $S$ versus the classical distance $\dH\xc$ for the 16 eligible crossings, with reference lines $S=\dH\xc$ and $S=2\dH\xc$. (d) Measured quantum failure versus $P_{\mathrm{LZ}}$ at $h=1$, $T=700/K$ for the instances with $P_{\mathrm{LZ}}\geq10^{-5}$ (10 weyl, 53 goto, out of 300 each); dashed line: equality. Open squares: the 40 goto instances whose minimum gap lies at $p=0$, the start of the ramp, where Eq.~\eqref{eq:LZ} does not apply. Dotted lines: mean failure of the instances with $P_{\mathrm{LZ}}<10^{-6}$ ($6.0\times10^{-5}$ weyl; $3.4\times10^{-4}$ for the goto instances with an interior minimum).}
\label{fig:law}
\end{figure*}

\section{From configuration crossings to quantum bottlenecks}
\label{sec:quantum}

Having isolated residual failures after matching initial fluctuations, we now test their spectral origin in the quantum KPO network. We first illustrate and test the alignment between classical crossings and quantum minimum-gap positions as $h$ decreases (Sec.~IV\,A), then examine finite-range exponential gap suppression (Sec.~IV\,B), and finally validate the resulting Landau--Zener predictions against direct four-spin quantum propagation (Sec.~IV\,C).

\subsection{A resolved example and the ensemble correspondence}
Instance \#29 illustrates two preexisting configuration branches exchanging energy order. Figure~\ref{fig:anatomy}(a) plots their classical energies relative to the Ising ground-configuration branch: the competing and ground branches are first resolved at approximately $p=0.150$ and $0.250$, respectively, and exchange energy order at $\pc=1.625$, with separation diagnostics $\dH=2$ and $\xc=1.418$.

Figure~\ref{fig:anatomy}(b) compares this classical crossing against the lowest even-parity quantum gap of $\hat H$. At $h=1$, the minimum gap occurs at $\ps=4.273$ with $\Dmin/K=0.02692$; reducing $h$ to $0.25$ shifts the minimum to $\ps=1.974$ and suppresses the gap to $\Dmin/K=1.265\times10^{-6}$. Decreasing $h$ thus draws the quantum bottleneck toward $\pc$ while strongly narrowing the gap.

Across the 60-instance four-spin spectral sample (all 17 test crossings and 43 noncrossing controls), Fig.~\ref{fig:law}(a) compares the minimum-gap position at $h=1$ and extrapolated to $h\to0$ with $\pc$ for crossings and with $p_{\mathrm{first}}$ (the first numerically resolved branch onset) for controls. For each instance, we perform a quadratic extrapolation $\ps(h)=p_0+c_1h+c_2h^2$ across all seven sampled values $h\in\{1,0.7,0.5,0.35,0.25,0.18,0.12\}$ (all returned minima lie inside the scan interval). Unlike the gap-amplitude fits below, this position extrapolation retains $\ps$ values at small $h$ where $\Dmin$ falls below the $10^{-6}K$ amplitude-resolution floor; the accuracy of minimum-gap locations associated with unresolved gaps has not been independently quantified.

For the 17 crossings, the median extrapolated offset $p_0-\pc$ is $-0.038$, with 71\% within $\pm 0.2$ and a Spearman correlation of $0.975$ between $p_0$ and $\pc$; for the 43 controls, the median offset $p_0-p_{\mathrm{first}}$ is $-0.002$. These extrapolations support a close semiclassical correspondence with the classical branch structure. At $h=1$, however, the median crossing offset is $+3.29$, so locating the finite-$h$ bottleneck still requires the quantum spectrum.

\subsection{Finite-range tunneling signatures}
Coupling between competing configurations opens an avoided crossing that mediates transfer between branches while limiting adiabaticity under a finite-time ramp. Exponentially small coherent-state overlap motivates the tunneling ansatz $\Dmin/K\sim\exp(-S/h)$ with geometric scale $\dH\xc$, where $S$ is an effective action over the sampled range. Unequal mode amplitudes, the multidimensional tunneling path, and detuning-dependent interference oscillations (known even in a single parametrically driven oscillator~\cite{Marthaler2007PRA}) can introduce instance-specific deviations from a universal proportionality.

Figure~\ref{fig:law}(b) shows the gaps against $1/h$. We compare two-parameter exponential fits $\ln(\Dmin/K)=a-S/h$ against power-law fits $\ln(\Dmin/K)=b+\nu\ln h$ on identical resolved points ($h\leq0.5$ and $\Dmin>10^{-6}K$, requiring at least three eligible points per instance) by residual sum of squares in log-gap space. Sixteen of the 17 crossings qualify (instance \#267 has too few resolved points). Over this resolved finite-$h$ range, the exponential form yields the smaller residual sum of squares in 12 of the 16 crossings, compared with only 3 of the 43 controls.

Across the 16 eligible crossings, the fitted effective action $S$ correlates strongly with $\dH\xc$ [Spearman coefficient $0.944$; Fig.~\ref{fig:law}(c)], although the ratio $S/(\dH\xc)$ spans $0.17$--$1.40$ with median $0.55$ (note that instances \#6, \#29, \#94, and \#231 each contribute three eligible points). Comparison with the heuristic scales $S=\dH\xc$ and $S=2\dH\xc$ (coherent-state overlap) confirms substantial instance dependence alongside the strong rank correlation. Full gap data, including unresolved entries and instance \#267, are listed in Table~\ref{tab:crossing}.

\subsection{Principal dynamical failures and residuals}
Returning to the full 300-instance four-spin test set at $h=1$ and $T=700/K$, Fig.~\ref{fig:law}(d) compares the Landau--Zener estimate~\eqref{eq:LZ} against the final failure probability from direct quantum propagation for both the weyl and goto families. Instance \#267 is included here, since its $h=1$ spectrum and time evolution are fully resolved.

For the weyl family, the four instances with $P_{\mathrm{LZ}}>0.01$ match direct propagation closely (predicted/observed: \#6: $0.017/0.016$; \#29: $0.179/0.184$; \#231: $0.022/0.027$; \#267: $0.999/0.999$). For \#6 the agreement holds at $T=200/K$ (predicted/observed $0.313/0.314$) and $T=700/K$, but breaks down at slower ramps. Its crossing lies at $\ps=6.46$, close to $p_f$, and the ground state of $\hat H(p_f)$ itself fails the sign read-out with probability $0.012$; this is the failure that remains for a slow ramp ($0.012$ at $T=2000/K$, where $P_{\mathrm{LZ}}=9\times10^{-6}$). Across all 17 crossing instances, the mean LZ-predicted success is $0.928$, compared with $0.927$ from direct propagation, confirming that the resolved avoided crossing accounts for the principal finite-time quantum failures under this schedule.

The failures that remain where $P_{\mathrm{LZ}}$ is negligible have a different origin in the two families (ramps of $T=200/K$ and $2000/K$ were also run, on all 300 test instances in the weyl family and on instances 0--99 in the goto family). For the 289 weyl instances with $P_{\mathrm{LZ}}<10^{-6}$ the mean failure is $6.0\times10^{-5}$ at $T=700/K$ and $5.9\times10^{-5}$ at $T=2000/K$, with every instance within $10\%$ of its $T=700/K$ value; raising $n_{\max}$ from 22 to 26 (instances 0--9) changes it by at most $1.2\%$. In the tested instances, it is therefore neither a nonadiabatic transition nor a truncation error: diagonalizing $\hat H(p_f)$ for ten instances and reading out its ground state reproduces the $T=2000/K$ failure to $1\%$, indicating that final-state sign read-out accounts for most of the slow-ramp residual failure. At $p_f=7K$ the ground state is not fully confined to the sign patterns of the Ising ground configuration, and this sets the floor of the measured failure in Fig.~\ref{fig:law}(d).

In the goto family 70 of the 300 instances have their minimum gap at $p=0$ ($0.12K$--$0.33K$), where the linear ramp starts at a finite rate and Eq.~\eqref{eq:LZ} does not apply [open squares in Fig.~\ref{fig:law}(d)]. On 52 of them the failure at $T=700/K$ is $0.002$--$0.057$, and among the 21 of these instances with paired $T=700/K$ and $2000/K$ ramps the failure drops by a median factor of $8.1$ (against $(2000/700)^2=8.2$), consistent with the $T^{-2}$ trend expected for an excitation at the abrupt start of a linear ramp; on the other 18 it does not exceed $10^{-5}$. The goto instances with an interior minimum follow Eq.~\eqref{eq:LZ} where $P_{\mathrm{LZ}}$ is appreciable (for example \#1: $0.069/0.072$; \#94: $0.037/0.035$; \#231: $0.257/0.237$; \#6: $0.168/0.155$), and where $P_{\mathrm{LZ}}<10^{-6}$ (214 instances) their mean failure of $3.4\times10^{-4}$ shows the same $T^{-2}$ trend (a median factor of $7.7$ on the 65 instances with paired ramps). The weyl family, whose gap at $p=0$ is larger by about $2Kh$, shows no such contribution at $T\geq700/K$.

To test the crossing mechanism directly, Sec.~\ref{sec:intervention} alters the oscillator detunings while keeping the Ising couplings fixed.

\begin{table*}[t]
\caption{Composition of the 123-instance paired spectral sample drawn from the 1000-instance G4 ensemble. Branch labels follow Sec.~\ref{sec:branches_method}. Panels (a,b) of Fig.~\ref{fig:intervention} show the first four groups (119 instances); panel (c) compares classical success for KO and MO (66 instances). NGT comprises instances \#50, \#602, \#640, and \#648.}
\label{tab:intervention_groups}
\begin{ruledtabular}
\begin{tabular}{llcl}
Group & Branch classification P $\to$ U & $N$ & Figure~\ref{fig:intervention} \\
KO & Crossing $\to$ no crossing & 40 & (a,b,c) \\
MO & No crossing $\to$ crossing & 26 & (a,b,c) \\
BOTH & Crossing $\to$ crossing & 13 & (a,b) \\
CTRL & No crossing $\to$ no crossing & 40 & (a,b) \\
NGT & At least one setting is NGT & 4 & Listed in Appendix~\ref{app:intervention} \\
\end{tabular}
\end{ruledtabular}
\end{table*}

\begin{figure*}[t]
\centering\includegraphics[width=0.95\textwidth]{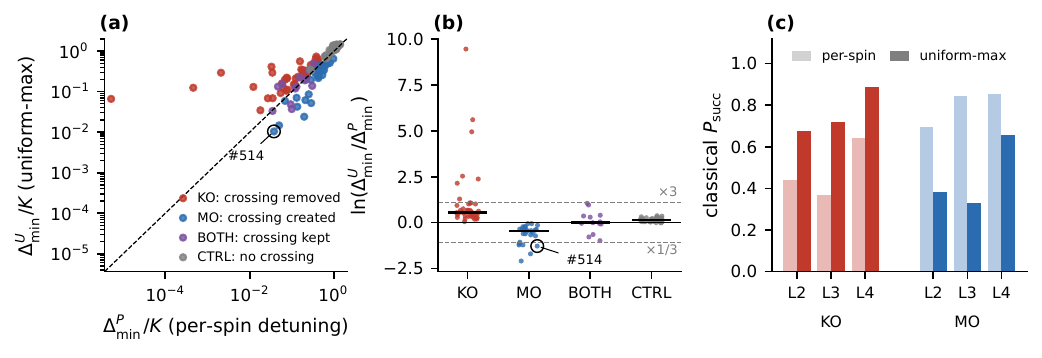}
\caption{Paired per-spin (P) and uniform-max (U) detuning intervention at fixed Ising couplings. (a,b) weyl-family minimum gaps at $h=1$ over $0\leq p\leq7$ for KO ($N=40$), MO ($N=26$), BOTH ($N=13$), and CTRL ($N=40$); the four NGT cases are listed in Appendix~\ref{app:intervention}. (a) Paired minimum gaps with dashed equality diagonal; MO instance \#514 has its U minimum at the upper scan boundary. (b) Per-instance log gap ratios $\ln(\Dmin^U/\Dmin^P)$ with group medians (black bars) and horizontal jitter; dashed horizontal lines mark reference ratios of $3$ and $1/3$ ($\pm\ln 3$). (c) Group-averaged classical success of L2, L3, and L4 ($\varepsilon=10^{-6}$, $T=700/K$, $R=1000$) for KO and MO before and after the detuning change.}
\label{fig:intervention}
\end{figure*}

\section{Paired detuning intervention}
\label{sec:intervention}
To test whether altering the classical configuration landscape shifts the quantum bottleneck directly, we compare per-spin and uniform-max detunings at fixed Ising couplings, tracking both the spectral gap response and classical success probabilities.

For each instance, the intervention replaces the reference detunings $\Delta_i$ by $D_{\max}=\max_i\Delta_i$ while keeping $J$, $K$, $\xi_0$, and $p_f$ fixed; superscripts P and U denote the per-spin and uniform-max settings, respectively. Because $J$ is unchanged, the discrete Ising objective and ground states remain identical, whereas the continuous oscillator landscape changes. The initial quadratic matrices satisfy

\begin{equation*}
\begin{aligned}
M_P&=\operatorname{diag}(\Delta_i)-\xi_0J,\\
M_U&=D_{\max}I-\xi_0J\\
   &=M_P+\operatorname{diag}(D_{\max}-\Delta_i).
\end{aligned}
\end{equation*}

Because $M_P\succeq0$ and $D_{\max}-\Delta_i\ge 0$, we retain $M_U\succeq0$, preserving the vacuum-ground-state condition of the weyl family (Sec.~\ref{sec:methods}). Unlike earlier KPO protocols that tune detuning to prepare the vacuum as an effective excited state~\cite{GotoKanao2020CommunPhys}, here detuning reshapes the ground-sector configuration branches; it also differs from the uniform mean $\bar\Delta$ used in the classical algorithmic study of Sec.~\ref{sec:sb}.

Across all 1000 G4 instances, switching from P to U removes 40 of the original 54 crossings, retains 13, and turns one (\#640) into an NGT case where the Ising ground configuration is no longer the lowest branch at $p_f=7$; another 26 instances acquire crossings. Our 123-instance paired spectral sample comprises all changed or retained crossing groups, the four NGT cases, and 40 noncrossing controls (the first 40 eligible instances in index order within 0--299), as summarized in Table~\ref{tab:intervention_groups}.

We evaluate the weyl-family spectra at $h=1$ over $0\leq p\leq7$, define the per-instance gap ratio $r_i=\Delta_{\min,i}^{U}/\Delta_{\min,i}^{P}$, and summarize each group by its geometric median $r_{\mathrm{typ}}=\exp[\operatorname{median}_i(\ln r_i)]$. Removing a crossing widens the minimum gap in all 40 KO instances ($r_{\mathrm{typ}}=1.733$), whereas inducing a crossing narrows it in all 26 MO instances ($r_{\mathrm{typ}}=0.632$) [Fig.~\ref{fig:intervention}(a,b)]; for one MO instance (\#514), the uniform-max minimum lies at the upper scan boundary $p=7$. By comparison, BOTH and CTRL show much smaller shifts ($r_{\mathrm{typ}}=0.991$ and $1.158$, respectively), reflecting mild background spectral changes when the crossing status is unchanged.

\begin{figure*}[t]
\centering\includegraphics[width=0.95\textwidth]{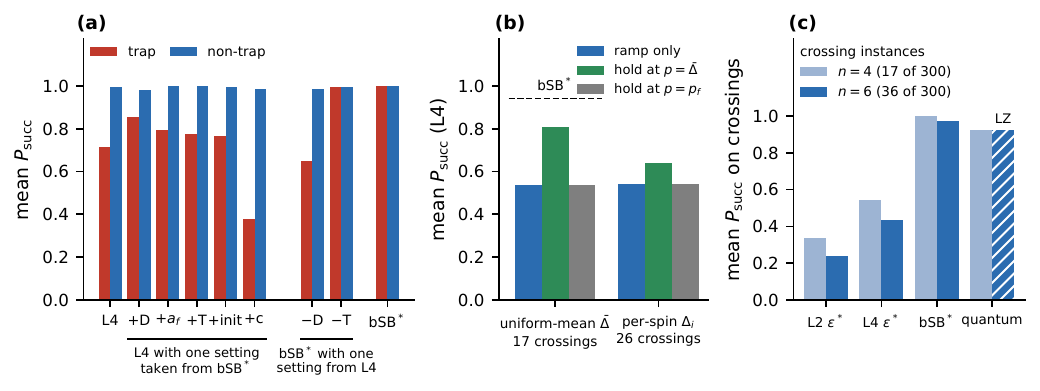}
\caption{Algorithmic ablations, Ising-window holds, and six-spin transfer. (a) Independent parameter replacements between L4 and \bSB{} on G4 0--499 ($R=1000$), grouped by the original per-spin trap label (43 trap, 457 non-trap). Plus signs replace one L4 setting with its \bSB{} value; minus signs restore the corresponding L4 setting in \bSB{}; init changes initialization amplitude and wall position jointly. (b) Equal-duration L4 hold tests ($\tau=1000/K$ added to a $700/K$ ramp) conditioned on each detuning's crossing subset (17 uniform-mean, 26 per-spin; five shared); the dashed \bSB{} line applies to the 17 uniform-mean crossings. (c) Mean success on the crossing instances at four and six spins (17 and 36 of the 300 instances of each size), with the classical parameters transferred without retuning. The four-spin quantum value is from direct propagation, the six-spin one (hatched, LZ) is the Landau--Zener estimate.}
\label{fig:sb}
\end{figure*}

To check the corresponding classical response, we run L2, L3, and L4 under the baseline protocol of Sec.~\ref{sec:methods} ($\varepsilon=10^{-6}$, $T=700/K$, $p_f=7$, $\delta t=0.0025/K$, $R=1000$). Figure~\ref{fig:intervention}(c) compares group-averaged success probabilities within KO and MO: in KO, mean L2/L3/L4 success rises from $0.438/0.366/0.642$ to $0.673/0.720/0.889$, whereas in MO it drops from $0.696/0.842/0.852$ to $0.381/0.330/0.658$. Classical success thus shifts in the same direction as the quantum gap across all three layers.

Taken together, the paired detuning intervention provides consistent directional support for the crossing-based interpretation: all 40 KO instances exhibit wider scanned minimum gaps and all 26 MO instances exhibit narrower gaps, while group-averaged classical success shifts in the corresponding direction for L2--L4. Because uniform detuning also shifts the overall energy scale (as reflected by the mild CTRL offset) and direct quantum time propagation is not performed under the modified setting, this test establishes a consistent directional link, though not a single-factor quantitative decomposition. Section~\ref{sec:sb} next examines how modifying the classical dynamical rules themselves provides a complementary route to high optimization performance.

\section{Classical alternatives and the limits of a shared origin}
\label{sec:sb}
While Sec.~\ref{sec:intervention} provided directional support for the crossing-based interpretation within the KPO setting, classical SB algorithms can alter both the dynamical rules and parameter normalizations. Below, we first decompose key parameter changes between L4 and \bSB{} (Sec.~VI\,A), then test the proposed uniform-detuning Ising window via equal-duration holds (Sec.~VI\,B), and finally examine how these findings transfer to six-spin instances (Sec.~VI\,C).

\subsection{Algorithmic changes beyond the KPO reference}
On the 300-instance four-spin test set, \bSB{} reaches a mean success probability of $1.000$, removing the quantum success-rate advantage observed against the KPO-parameter classical layers under the stated protocols of Sec.~\ref{sec:methods}.

Figure~\ref{fig:sb}(a) probes individual parameter replacements between L4 and \bSB{} on the 500-instance mechanism set (G4 indices 0--499), partitioned by the original per-spin trap label (43 trap and 457 non-trap instances; $R=1000$ trajectories per instance). Starting from L4, $+$D, $+a_f$, $+$T, and $+$c independently substitute the \bSB{} detuning, final pump, duration, and coupling normalization, while $+$init jointly replaces the initialization amplitude and wall position (only their ratio matters; Appendix~\ref{app:ablation}); conversely, $-$D and $-$T start from \bSB{} and restore the corresponding L4 settings (full parameter mappings and additional single-parameter ablations appear in Appendix~\ref{app:ablation}). On this mechanism set, \bSB{} averages $0.998$ overall and $1.000$ on the trap subset.

Uniform-mean detuning ($+$D) raises L4 trap-subset success from $0.714$ to $0.856$, whereas restoring per-spin detuning in \bSB{} ($-$D) drops it to $0.651$, identifying detuning uniformity as a major factor. By contrast, substituting only the coupling normalization ($+$c) reduces L4 trap success to $0.377$, whereas restoring the shorter L4 duration in \bSB{} ($-$T) leaves trap success at $0.997$. Because no single replacement raises L4 to the \bSB{} level, the algorithmic advantage reflects coupled parameter choices rather than any single isolated ingredient.

Other SB extensions, such as deterministic heating inspired by Nos\'{e}--Hoover dynamics~\cite{Kanao2022CommunPhys} and replica-coupled SB~\cite{Pawlowski2026arXiv}, further broaden the classical design space; here we focus on the tuned \bSB{} and \dSB{} baselines of Sec.~\ref{sec:methods}.

\subsection{Testing the Ising window}
Uniform detuning creates a pump region where coupling forces can reorganize boundary configurations. Inside the walls, the L4 momentum equation is

\begin{equation*}\begin{aligned}
\dot y_i&=-(D-p)x_i+\xi_0\sum_jJ_{ij}x_j,\\
\dot y_i\big|_{p=D}&=\xi_0\sum_jJ_{ij}x_j.
\end{aligned}\end{equation*}

At $p=D$, the local restoring term vanishes simultaneously for all oscillators. At a corner $x_i=w s_i$, any spin with $s_i\sum_jJ_{ij}s_j<0$ experiences an inward coupling force, directly linking corner instability to an energy-lowering Ising flip. Here $D=\bar\Delta$ is the uniform mean used in the classical algorithm (distinct from $D_{\max}$ in Sec.~\ref{sec:intervention}, which preserves the quantum vacuum condition). For per-spin detunings, holding at $p=\bar\Delta$ serves as a matched control in which the local terms do not vanish simultaneously.

On the same 500-instance mechanism set (L4 initialization, walls, $\delta t=0.0025/K$, $R=1000$), we insert a hold of duration $\tau=1000/K$ into the $T=700/K$ linear ramp to $p_f=7$, either at $p=\bar\Delta$ or at the final pump $p_f$. Both schedules therefore share the same total duration $1700/K$, isolating where the additional evolution time is spent.

Figure~\ref{fig:sb}(b) evaluates each detuning setting on its own crossing subset: 17 instances for uniform-mean detuning and 26 for per-spin detuning (five shared). For uniform-mean detuning, holding at $p=\bar\Delta$ raises mean crossing success from $0.535$ to $0.806$ (a gain of $+0.271$), whereas an equal-duration hold at $p_f$ leaves success unchanged at $0.535$. Under per-spin detuning, the corresponding values are $0.541$, $0.641$, and $0.541$.

This equal-duration comparison directly confirms that additional evolution is effective specifically within the uniform-detuning Ising window rather than at the ramp endpoint. At the same time, the window-hold mean ($0.806$) remains below the \bSB{} reference ($0.941$) on the same 17 uniform-mean crossings, showing that holding inside the uniform-detuning Ising window substantially improves L4 success and narrows the performance gap to \bSB{} on the same subset, while leaving part of the gap unresolved.

\subsection{Six-spin observations}
To test transfer to a larger network, we apply the four-spin-selected hyperparameters to all 300 G6 instances without retuning. Figure~\ref{fig:sb}(c) compares the mean success on the crossing instances of the two sizes. Crossings are roughly twice as prevalent at six spins: 36 of 300 instances ($12.0\%$, with six additional NGT cases kept in the denominator), compared with 17 of 300 ($5.7\%$) at four spins. On the 36 six-spin crossings, tuned L2, L4, and \bSB{} achieve mean successes of $0.239$, $0.432$, and $0.972$, respectively (under the $T=700/K$, $R=1000$ KPO protocol and the $10^4$-step \bSB{} budget).

Quantum spectra at $h=1$ cover 66 six-spin instances (all 36 crossings and 30 noncrossing controls, with extended pump scans used for nine instances whose initial minima lay at the scan boundary). Their median minimum gaps are $0.326K$ for crossings versus $0.914K$ for noncrossings. Estimating quantum success on the 36 crossings via $P_Q^{\mathrm{LZ}}=1-P_{\mathrm{LZ}}$ from Eq.~\eqref{eq:LZ} at $T=700/K$ gives a mean of $0.927$. Unlike the four-spin bar in Fig.~\ref{fig:sb}(c) (obtained by direct propagation), this six-spin quantum estimate relies on the isolated two-level assumptions of Sec.~\ref{sec:quantum_method}.

Ten six-spin instances exhibit zero observed \bSB{} success across all sampled trajectories: eight lie outside both the trap and crossing classes, while \#215 is NGT and \#243 is a crossing (only \#243 lies in the spectral sample, and direct six-spin quantum propagation is unavailable). These cases highlight failure modes beyond our current structural classification. Moreover, because the coupling generator is not normalized with system size (raising the mean base detuning from approximately $0.744$ to $1.254$), these six-spin results serve as a transferred-parameter check at a second system size rather than a finite-size scaling law.

\section{Discussion and conclusion}
\label{sec:discussion}
Classical configuration crossings organize a distinct class of quantum bottlenecks in the investigated KPO networks. Extrapolated minimum-gap positions, finite-range exponential gap suppression, and instance-level agreement between Landau--Zener estimates and direct quantum evolution connect branch competition directly to finite-time quantum failures on the same instances. Paired detuning interventions confirm the direction of this link: minimum gaps shift consistently across all affected crossing instances, accompanied by matching shifts in group-averaged classical success. Together with the matched-fluctuation TWA controls, this diagnosis moves beyond aggregate quantum-versus-classical comparisons by isolating the structural origin of residual difficulty.

A shared quantum origin does not, however, fix the performance of derived classical algorithms. Tuned \bSB{} eliminates the quantum success-rate advantage on the four-spin test set by modifying the dynamical rules rather than shadowing the classical KPO flow, while equal-duration hold tests show that evolving inside a uniform-detuning Ising window substantially improves L4 success and narrows the gap to \bSB{} on the matched subset. On the quantum side, excited-state KPO protocols likewise demonstrate that ground-state adiabatic passage is not the only available route~\cite{GotoKanao2020CommunPhys}. Quantum simulations of dissipative coherent Ising machines also link improved optimization to escape from false minima~\cite{Thenabadu2025arXiv}, although their open-system dissipation, state preparation, and time-to-solution criteria differ from our closed, vacuum-initialized, final-time setting.

Recent superconducting-circuit experiments provide natural platforms for probing these spectral and dynamical questions: pairwise level degeneracies have been resolved in a double-well Kerr parametric oscillator~\cite{Frattini2024PRX}, and frequency-chirped drives have been demonstrated in a two-KPO annealing device~\cite{Yamaji2025PRApplied}. Extending the present closed-system, fixed-ramp diagnosis to such hardware will require incorporating single- and two-photon loss together with dephasing.

Several physical boundaries define the scope of our findings. Classical trajectories continue to switch sign patterns well after initial amplification, so a branch diagram is a structural diagnostic of the energy landscape rather than a literal capture history. At $h=1$, minimum-gap positions can lie far from $\pc$, the fitted effective action $S$ exhibits substantial instance dependence relative to $\dH\xc$, and the failure does not vanish for slow ramps, because the ground state at $p_f=7K$ is not fully confined to the Ising ground configuration (most visibly when the crossing lies close to $p_f$, as for \#6). Moreover, while our results are consistent with tunneling across competing configurations, they do not by themselves establish coherence necessity for every recovered instance; six-spin quantum success is estimated via LZ rates rather than direct propagation, and exhaustive enumeration of $2^{n-1}$ branches does not scale to large networks.

Within these boundaries, revisiting the quantum origins of simulated bifurcation clarifies a key distinction: tunneling bottlenecks govern finite-time failures in the parent KPO model, whereas classical SB algorithms achieve high optimization performance by altering the dynamics used to explore the same Ising objective. How tailored finite-time schedules exploit this crossing structure, and how photon loss reshapes the resulting output distribution, remain important directions for future control studies.
\begin{acknowledgments}
Y.C.\ acknowledges a postdoctoral fellowship from the S\~ao Paulo Research Foundation (FAPESP), process 2025/25586-5. This study was financed, in part, by the S\~ao Paulo Research Foundation (FAPESP), Brasil. Process Numbers \#2025/25586-5 and \#2021/02120-0. The computations were performed on a server with eight NVIDIA RTX PRO 6000 GPUs. Generative AI tools (OpenAI ChatGPT, Anthropic Claude and Google Gemini) were used to assist with software development, and ChatGPT for language editing; the authors reviewed and verified all AI-assisted output and are responsible for the manuscript and the software.
\end{acknowledgments}
\section*{Code and data availability}
The implementation, run instructions, instance sets, per-instance result files, and the scripts for all figures are available at \url{https://github.com/yaochengchen/sb-quantum-origins}. The code is released under the MIT License and the instance sets, results, and figures under CC BY 4.0.
\section*{Conflict of interest}
The authors declare no conflict of interest.
\appendix
\section{Numerical verification}
\label{app:conv}
The Fock-space occupation cutoff is $n_{\max}$ per mode (local dimension $n_{\max}+1$). Across the 300 weyl-family $T=700/K$ propagation runs, the maximum final-state combined population in the two highest retained Fock levels (maximized over modes) is $2.3\times10^{-6}$ at $h=1$ and $2.0\times10^{-5}$ at $h=0.35$; across all reported $h=1$ goto-family propagation runs ($T=700/K$ and $T=200/K$), it remains below $2.0\times10^{-5}$.

Paired cutoff-extension checks were carried out over G4 instances 0--9 (including the crossing instances \#1 and \#6). At $h=1$, increasing $n_{\max}$ from $22$ to $26$ changes $P_{\mathrm{succ}}$ by at most $4.0\times10^{-5}$ in the weyl family (at \#6; mean $4.3\times10^{-6}$, with all other nine instances below $4.5\times10^{-7}$) and at most $3.4\times10^{-4}$ in the goto family (at \#6; mean $3.5\times10^{-5}$). This extension reduces the maximum top-two-level population below $1.4\times10^{-8}$ (weyl) and $8.4\times10^{-8}$ (goto). At $h=0.5$ in the weyl family, raising $n_{\max}$ from $32$ to $36$ over the same 10 instances yields $\max|\Delta P_{\mathrm{succ}}|=2.6\times10^{-4}$ (at \#6; mean $2.7\times10^{-5}$) and reduces the maximum top-two-level population from $1.2\times10^{-5}$ to $3.2\times10^{-7}$.

Verification checks compare sparse and tensor Hamiltonian actions, fixed-pump matrix exponentials, and time-dependent propagation. With the step cap $0.1$ used in all propagation runs, a $T=700/K$ ramp on a two-oscillator test system agrees with DOP853 to $7\times10^{-6}$ in the state amplitudes ($1.4\times10^{-8}$ with cap $0.05$); in the four-spin runs the step is set by the Lanczos error estimate (tolerance $10^{-8}$ for the whole ramp) and averages $0.047/K$, below the cap, alongside CPU/GPU consistency checks on small test systems. For the classical KPO-parameter layers L2--L4, paired step-size checks were performed on G4 instances 0--9 ($R=1000$). Halving $\delta t$ from $0.0025$ to $0.00125$ changes $P_{\mathrm{succ}}$ by at most $0.006$ per instance (L3; $0.001$ for L2 and L4) and at most $6\times10^{-4}$ in the 10-instance mean (L3; $1\times10^{-4}$ for L2 and L4). Doubling $\delta t$ to $0.005$ likewise yields differences at most $0.005$ per instance and $5\times10^{-4}$ in the mean. For the TWA (L1) single instances change by up to $0.04$ under either step change, without decreasing on refinement, which reflects the sensitivity of individual trajectories rather than integration error; the 10-instance mean changes by at most $0.004$.

\section{Spectral resolution and fitting}
\label{app:gap}
The eigensolver tracks the lowest even-parity eigenpairs along a pump grid and refines candidate minima by bounded Brent search (the coarse grid alone can miss narrow avoided crossings), using a default residual tolerance of $10^{-8}$ (requiring eigenpair residuals $\le 10^{-7}\max(1,|E|)$, a relative criterion). Across all 60 four-spin spectral instances at $h\in\{0.7,0.5,0.35,0.25,0.18,0.12\}$, every run converges within tolerance with all minima strictly inside the scan interval. The maximum top-two-level ground-state population ranges from $1.4\times10^{-5}$ ($h=0.12$) to $7.9\times10^{-5}$ ($h=0.35$). Similarly, the 66 six-spin spectral runs at $h=1$ ($n_{\max}=18$) keep the maximum top-two-level population at $1.5\times10^{-5}$. Gaps below the $10^{-6}K$ resolution floor are excluded from the action fits, which require at least three eligible points per instance.
\begin{table*}[tbp]
\caption{The 17 crossing instances of the test set (weyl family): Hamming distance $\dH$ and mean squared amplitude $\xc$ at the classical crossing $\pc$; quantum gap position at $h=1$ and extrapolated to $h\to0$ (quadratic in $h$); minimum gap at four values of $h$. $\dagger$: below $10^{-6}$, at the eigensolver resolution, not used in fits.}
\label{tab:crossing}
\begin{ruledtabular}
\begin{tabular}{rcccccllll}
\# & $\dH$ & $\xc$ & $\pc$ & $\ps_{h=1}$ & $\ps_{h\to0}$ & $\Dmin^{h=1}$ & $\Dmin^{h=0.5}$ & $\Dmin^{h=0.25}$ & $\Dmin^{h=0.12}$\\
\hline
1   & 2 & 0.306 & 0.485 & 3.77 & 0.29 & 7.9e-2 & 5.2e-2 & 2.7e-2 & 5.9e-3\\
6   & 1 & 3.168 & 3.645 & 6.46 & 3.87 & 3.3e-2 & 3.3e-3 & 1.3e-6 & $\dagger$\\
18  & 1 & 0.363 & 0.740 & 3.71 & 0.70 & 4.2e-1 & 3.1e-1 & 2.5e-1 & 2.1e-1\\
29  & 2 & 1.418 & 1.625 & 4.27 & 1.71 & 2.7e-2 & 3.4e-3 & 1.3e-6 & $\dagger$\\
38  & 1 & 1.425 & 1.785 & 5.09 & 1.63 & 1.6e-1 & 8.3e-2 & 1.5e-2 & 6.3e-5\\
44  & 1 & 0.825 & 1.160 & 4.62 & 0.93 & 2.3e-1 & 1.6e-1 & 8.4e-2 & 1.4e-2\\
49  & 1 & 0.433 & 0.800 & 3.89 & 0.78 & 4.0e-1 & 3.2e-1 & 2.5e-1 & 1.7e-1\\
94  & 1 & 2.942 & 3.420 & 5.84 & 3.58 & 7.6e-2 & 6.6e-3 & 8.3e-6 & $\dagger$\\
99  & 2 & 0.183 & 0.610 & 4.19 & 0.54 & 1.1e-1 & 8.3e-2 & 5.6e-2 & 3.0e-2\\
133 & 1 & 0.956 & 1.215 & 5.40 & 0.84 & 7.7e-2 & 5.3e-2 & 2.6e-2 & 2.1e-3\\
134 & 2 & 0.050 & 0.215 & 3.60 & 0.21 & 1.2e-1 & 8.7e-2 & 6.0e-2 & 3.9e-2\\
162 & 2 & 0.076 & 0.760 & 4.66 & 0.83 & 7.4e-2 & 5.9e-2 & 4.6e-2 & 3.5e-2\\
190 & 1 & 0.524 & 0.640 & 5.16 & 0.34 & 5.4e-2 & 4.1e-2 & 2.6e-2 & 9.6e-3\\
218 & 1 & 0.903 & 1.265 & 5.40 & 0.96 & 9.9e-2 & 7.9e-2 & 4.6e-2 & 6.4e-3\\
231 & 2 & 1.587 & 1.965 & 4.53 & 2.04 & 5.8e-2 & 9.6e-3 & 1.3e-5 & $\dagger$\\
264 & 1 & 0.524 & 0.810 & 4.06 & 0.66 & 2.6e-1 & 1.7e-1 & 8.8e-2 & 2.0e-2\\
267 & 2 & 3.845 & 4.190 & 5.65 & 4.19 & 4.6e-4 & $\dagger$ & $\dagger$ & $\dagger$\\
\end{tabular}
\end{ruledtabular}
\end{table*}

\section{Sample definitions and classical hyperparameters}
\label{app:hyper}

Table~\ref{tab:samples} summarizes the instance ensembles, index ranges, and analysis subsets used throughout the paper.

\begin{table*}[tbp]
\caption{Instance ensembles, index ranges, and analysis subsets. IDs are local to G4 or G6. Spectral subsets are selected specifically for crossing--spectrum analysis rather than ensemble-wide averages.}
\label{tab:samples}
\begin{ruledtabular}
\begin{tabular}{p{0.19\textwidth}p{0.22\textwidth}p{0.25\textwidth}p{0.26\textwidth}}
Sample & Size and membership & Purpose & Notes \\
\hline
G4 ensemble & 1000; IDs 0--999 & Training, testing, and structural analyses & Parent four-spin ensemble. \\[3pt]
G4 performance test & 300; IDs 0--299 & Held-out evaluation & Disjoint from training; 17 crossings. \\[3pt]
Noise training & 200; IDs 300--499 & Select L2--L4 initial noise & Subset of SB training. \\[3pt]
SB training & 700; IDs 300--999 & Select bSB/dSB hyperparameters & Separate search space. \\[3pt]
G4 small-$h$ spectra & 60; 17 crossings + 43 controls & Crossing--spectrum relation & IDs 0--49 plus all test crossings. \\[3pt]
G6 ensemble & 300; IDs 0--299 & Six-spin transfer test & Evaluated without retuning. \\[3pt]
G6 spectra & 66; 36 crossings + 30 controls & Quantum LZ estimates & All crossings plus first 30 noncrossings. \\[3pt]
Other mechanism sets & Specified with each analysis & Noise and intervention analyses & May overlap training indices. \\[3pt]
\end{tabular}
\end{ruledtabular}
\end{table*}

The classical KPO layers were scanned over $\varepsilon\in\{10^{-8},10^{-6},10^{-4},10^{-3},10^{-2},0.03,0.1,0.3,1\}$. Maximizing the training-set mean selects $\varepsilon^*=0.03$ (L2), $10^{-3}$ (L3), and $10^{-6}$ (L4, which is nearly insensitive to $\varepsilon$). For bSB and dSB, we searched the grid $c\in\{0.25,0.5,0.7,1,1.5\}$, $\delta t\in\{0.1,0.25,0.5,1,1.25\}$, $N_{\mathrm{steps}}\in\{100,300,10^3,3\times10^3,10^4\}$, and $A\in\{10^{-6},10^{-3},0.1,1\}$ (initial $x,y$ uniform in $[-A,A]$, $R=1000$, using common random numbers across the grid). The selected configurations are listed in Table~\ref{tab:baselines}; several bSB configurations tie at the training-set optimum ($0.997$) and all achieve $1.000$ on the test set.

\subsection{Ablation settings and additional results}
\label{app:ablation}
\begin{table}[t]
\caption{Reference settings for the single-parameter ablations in units of $K=1$. Here $\bar\Delta$ and $\sigma_J$ are instance-specific; each replacement holds all other reference settings fixed.}
\label{tab:ablation}
\begin{ruledtabular}
\begin{tabular}{lcc}
Setting & L4 & \bSB{} in KPO units \\
\hline
$D$ & $\Delta_i$ & $\bar\Delta$ \\
$c$ & $\xi_0$ & $0.5\bar\Delta/(\sigma_J\sqrt n)$ \\
$a_f$ & $7$ & $\bar\Delta$ \\
$T$ & $700$ & $10^4/\bar\Delta$ \\
$\delta t$ & $0.0025$ & $1/\bar\Delta$ \\
init: $(A,w)$ & $(10^{-6},\sqrt7)$ & $(10^{-3},1)$ \\
\end{tabular}\end{ruledtabular}\end{table}

The ablation expresses \bSB{} in KPO time units (Table~\ref{tab:ablation}). Its unmodified reference uses $10^4$ steps; modified configurations use $N_{\mathrm{steps}}=\lceil T/\delta t\rceil$ and adjust the actual step to $T/N_{\mathrm{steps}}$. Initial $x$ and $y$ coordinates are independent uniform samples on $[-A,A]$, reusing the same random stream across configurations of each instance, and the init replacement changes $A$ and $w$ jointly. Because the L4 and \bSB{} equations are linear inside the walls, rescaling $x$, $y$, and $w$ together leaves every sign unchanged, so only the ratio $A/w$ matters and init is a single parameter ($A/w=3.8\times10^{-7}$ for L4 and $10^{-3}$ for \bSB{}). Additional single-parameter ablations not shown in Fig.~\ref{fig:sb}(a) (L4 $+\delta t$, L4 $+a_f$ at $2\bar\Delta$ and $4\bar\Delta$, and the reverse \bSB{} replacements $-c$, $-a_f$, $-\delta t$, and $-$init) yield trap-subset mean successes of $0.706$, $0.788$, $0.781$, $1.000$, $1.000$, $1.000$, and $1.000$, respectively.

\section{Exceptional finite-pump cases in the detuning intervention}
\label{app:intervention}
Four instances in the 123-instance paired detuning sample of Sec.~\ref{sec:intervention} are classified as NGT under either per-spin (P) or uniform-max (U) detuning, meaning that the Ising ground configuration is not the lowest classical branch at $p_f=7$. Their minimum gaps over the scanned pump interval $0\leq p\leq 7$ are listed in Table~\ref{tab:neg}; within this group, one P minimum and three U minima lie at the scan boundary.

For transparency with the archived analysis scripts, the original internal amplitude targets were a geometric-median gap ratio of at least $3$ for KO and at most $1/3$ for MO [the dashed lines in Fig.~\ref{fig:intervention}(b)], and a uniform-mean Ising-window hold success of at least $0.90$ in Sec.~\ref{sec:sb}. While the directional criteria (40/40 KO widening, 26/26 MO narrowing, and matching classical shifts) and the strong improvement of window holds over endpoint holds ($0.806$ versus $0.535$) are met, those stricter amplitude targets were not reached ($1.733$, $0.632$, and $0.806$, respectively), so the main text restricts its claims to directional support and substantial improvement over the endpoint control.

\begin{table}[t]
\caption{The four NGT cases in the paired detuning intervention sample. Minimum gaps are evaluated over $0\leq p\leq 7$ in units of $K$.}
\label{tab:neg}
\begin{ruledtabular}
\begin{tabular}{rccrr}
Instance & P class & U class & $\Delta^P_{\min}/K$ & $\Delta^U_{\min}/K$ \\
\hline
50 & No crossing & NGT & 0.208702 & 0.116497 \\
602 & NGT & No crossing & 0.032464 & 0.028264 \\
640 & Crossing & NGT & 0.058200 & 0.020386 \\
648 & No crossing & NGT & 0.019348 & 0.039303 \\
\end{tabular}
\end{ruledtabular}
\end{table}

\bibliography{references}
\end{document}